# Status of the Higgs Mass Bound[*]

Urs M. Heller[a]

[a]SCRI, The Florida State University, Tallahassee, FL 32306-4052, USA

The status of the triviality bound of the Higgs mass in the Minimal Standard Model is reviewed. It is emphasized that the bound is obtained, in the scalar sector, by limiting cutoff effects on physical processes. Results from several regularization schemes, including actions that allow a parameterization and tuning of the leading cutoff effects, are presented. They lead to the conclusion that the Minimal Standard Model will describe physics to an accuracy of a few percent up to energies of the order 2 to 4 times the Higgs mass, $M_H$, only if $M_H \leq 710 \pm 60\ GeV$. The status of Higgs and fermion mass bounds in Higgs-fermion models is also briefly reviewed.

## 1. Introduction

The idea that the triviality of the scalar sector of the Minimal Standard Model (MSM) leads to an upper bound on the Higgs mass [1] is by now well established. For recent reviews and references see e.g. [2,3]. The statement about triviality is embodied in the relation between the cutoff $\Lambda$, introduced to regularize the theory, the physical Higgs mass, $M_H$, and the renormalized coupling, $g_R = 3 M_H^2/F^2$ ($F = 246\ GeV$ is the weak scale), which at the two-loop level reads

$$\frac{M_H}{\Lambda} = C \left(\frac{g_R}{4\pi^2}\right)^{13/24} \exp\left\{-\frac{4\pi^2}{g_R}\right\} [1 + O(g_R)] \quad (1)$$

Here $C$ is a constant which depends on the bare parameters but is not computable in perturbation theory. From eq. (1) we see that the limit $\Lambda \to \infty$ implies $g_R \to 0$, i.e. we are left with a non-interacting, **trivial** theory. But we need an interacting scalar sector for the Higgs mechanism to work and give masses to the intermediate vector bosons $W$ and $Z$. Therefore we need to keep the cutoff finite: the MSM has to be viewed as an effective theory that describes physics at energies below the cutoff scale.

Since we have to keep a finite cutoff, we may ask what happens if we try to make the (renormalized) scalar self-interactions stronger. Eq. (1) tells us that as $g_R$ increases, so does the ratio $M_H/\Lambda$. But since the Higgs mass is one of the physical quantities that the MSM is supposed to describe, we certainly need $M_H/\Lambda < 1$. Hence we have arrived at an upper bound on the Higgs mass, the so called triviality bound.

Specifying more precisely the meaning of the "$<$" in the criterion $M_H/\Lambda < 1$, a rather accurate bound of $710 \pm 60\ GeV$ has been obtained in the pure scalar sector. I shall review this in section 2. Such a heavy Higgs is rather strongly interacting, with a width $\gtrsim 25\%$ of its mass. The renormalized coupling goes up to about $85\% - 100\%$ of the tree-level unitarity bound [4,5], depending on whether the Higgs contribution is counted in the unitarity balance or not. Hence we are reaching the limits of renormalized perturbation theory, though it seems to work rather well in the entire possible Higgs mass range.

At the scales relevant to the Higgs bound, the gauge couplings of the MSM are small and can be treated perturbatively. Concentrating on the pure scalar sector in obtaining the upper bound on the Higgs mass is then justified as long as the heaviest fermion with a Yukawa coupling to the Higgs, the top quark, is not heavier than about $200\ GeV$, as is favored by experiment [6]. Then all the Yukawa couplings can be treated perturbatively as well. However, it is conceivable that a fourth, as yet undetected family of heavy fermions exists, as long as its neutrino is heavier than half the $Z$ mass – the limit on the number of light neutrinos is $N_\nu = 3.04 \pm 0.04$ [6]. Therefore, and also for purely theoretical reasons – the Yukawa interaction is also (perturbatively) trivial – the ques-tion of bounds on Higgs and fermion masses have recently been studied in several Yukawa models. I shall give a brief review of the present status in section 3. Compared to the pure scalar case the studies so far are rather preliminary. In particular, due to the difficulties with chiral fermions on a lattice, a satisfactory, and numerically manageable, lattice transcription of the MSM is still missing. Nevertheless, no Yukawa couplings greatly exceeding the tree-level unitarity bounds have been found.

---

[*]To appear in the Proceedings of Lattice '93, October 12–16, 1993 Dallas TX.

## 2. Pure Scalar Sector

### 2.1. Cutoff Effects and Generalized Actions

The triviality of the scalar sector of the MSM has been established for some time both analytically [5,7] and numerically [2,8,9]. As a consequence, all observable predictions have some cutoff dependence, with the cutoff provided by some, as yet unknown, embedding theory. We are interested in the situation where for sufficiently small $M_H/\Lambda$ any process with energy $E \lesssim 4M_H$ has only small order $1/\Lambda^2$ cutoff corrections. Then, for energies up to about $4M_H$, the scalar sector is representable by an effective action [10,3]

$$L_{\text{eff}} = L_{\text{ren}} + \frac{1}{\Lambda^2} \sum_A c_A O_A , \quad \dim O_A \le 6 \quad (2)$$

with $L_{\text{ren}}$ the usual renormalized $\phi^4$ Lagrangian. $O_A$ are operators with the correct symmetry properties and dimension less than or equal to 6, and the coefficients $c_A$ depend on the embedding theory. Eliminating redundant operators, which leave the S-matrix unchanged, there are two "measurable" $c_A$'s.

The restriction on cutoff effects translates into some limits on the $c_A$'s. We don't know the true embedding theory, but almost any reasonable cutoff model with enough free parameters can produce the same $M_H$ and same $c_A$'s as the true theory on the level of $L_{\text{eff}}$, eq. (2). Thus, if we look at a large enough class of such cutoff models we will likely find the bound on the Higgs mass that is obeyed by the "true" Higgs particle.

The most straightforward implementation is to start with an action of the form as in eq. (2) on the level of bare fields and parameters. It turns out, however, that the maximal renormalized coupling $g_R$ is obtained at maximal bare $\phi^4$ self coupling, i.e., in the nonlinear limit (see e.g. [5,11]), where all dimension six operators become trivial. In a non-linear theory it is terms with four derivatives that allow us to tune the cutoff effects of order $1/\Lambda^2$. Maintaining $O(N)$ invariance – a useful generalization of the $O(4)$ invariance –, there are three different terms with four derivatives, and we are led to consider actions of the form

$$S = \int_x \left[ \frac{1}{2} \vec{\phi}_c (-\partial^2 + 2b_0 \partial^4) \vec{\phi}_c - \frac{b_1}{2N} (\partial_\mu \vec{\phi}_c \cdot \partial_\mu \vec{\phi}_c)^2 \right.$$
$$\left. - \frac{b_2}{2N} (\partial_\mu \vec{\phi}_c \cdot \partial_\nu \vec{\phi}_c - \frac{1}{4} \delta_{\mu,\nu} \partial_\sigma \vec{\phi}_c \cdot \partial_\sigma \vec{\phi}_c)^2 \right] \quad (3)$$

with $\vec{\phi}_c^2 = N\beta$ fixed. Up to terms with more derivatives, the parameter $b_0$ can be eliminated with a field redefinition

$$\vec{\phi}_c \to \frac{\vec{\phi}_c + b_0 \partial^2 \vec{\phi}_c}{\sqrt{\vec{\phi}_c^2 + b_0^2 (\partial^2 \vec{\phi}_c)^2 + 2b_0 \vec{\phi}_c \partial^2 \vec{\phi}_c}} \sqrt{N\beta}. \quad (4)$$

This leaves two free parameters to tune the cutoff effects, exactly the number of measurable coefficients $c_A$ in eq. (2). Therefore eq. (3) should give a good parameterization to obtain the Higgs mass bound.

The earlier studies of the Higgs mass bound, using non-linear nearest neighbor actions on hypercubic[2] [5,8] or $F_4$ lattices [9] as well as a Symanzik improved action [12] were just special cases. A posteriori it turned out that they came quite close to the actual bound.

### 2.2. Solution at large $N$

Before attempting numerical studies of lattice transcriptions of the class of actions, eq. (3), one would like to have some intuitive understanding of the effects of the four-derivative couplings. To this end these models were analyzed analytically in the soluable large $N$ limit [11]. Considering a class of Pauli-Villars and several lattice regularizations of action eq. (3), it was found that, after $b_0$ has been eliminated, $b_2$ has no effect and that the bound depends monotonically on $b_1$, increasing with decreasing $b_1$. Overall stability of the homogeneous broken phase restricts the range of $b_1$ and thus we find a finite optimal value for $b_1$. The physical picture that emerges is that among the nonlinear actions the bound is further increased by reducing as much as possible the attraction between low momentum pions in the $I = J = 0$ channel.

---

[2]For the hypercubic action, instead of the term proportional to $b_0$ there actually appears a Lorentz invariance breaking term of the from $\vec{\phi}_c \sum_\mu \partial_\mu^4 \vec{\phi}_c$.

When considering lattice regularizations of actions eq. (3), because we desire to preserve Lorentz invariance to order $1/\Lambda^2$, we use the $F_4$ lattice [13]. We started with the naïve nearest-neighbor model. Next we considered the simplest action that has a tunable parameter $b_1$. We should emphasize that on the $F_4$ lattice, unlike on the hypercubic lattice, this can be done in a way that maintains the nearest–neighbor character of the action, namely by coupling fields sited at the vertices of elementary bond-triangles. Because the dependence of physical observables on the bare action is highly nonlinear it turned out that the $O(p^4)$ part of the free inverse Euclidean propagator can influence the bound somewhat. So we added a term to eliminate the "wrong sign" order $p^4$ term in the free propagator, amounting to Symanzik improvement. These three $F_4$ actions are[3]

$$S_1 = -2\beta_0 \sum_{<x,x'>} \vec{\phi}(x) \cdot \vec{\phi}(x')$$
$$S_2 = S_1 - \frac{\beta_2}{8} S_{4\phi} \quad (5)$$
$$S_3 = -2(2\beta_0 + \beta_2) \sum_{<x,x'>} \vec{\phi}(x) \cdot \vec{\phi}(x')$$
$$+ (\beta_0 + \beta_2) \sum_{\ll x,x' \gg} \vec{\phi}(x) \cdot \vec{\phi}(x') - \frac{\beta_2}{8} S_{4\phi}$$

where

$$S_{4\phi} = \sum_x \sum_{\Delta_x} \left[ \left( \vec{\phi}(x) \cdot \vec{\phi}(x') \right) \left( \vec{\phi}(x) \cdot \vec{\phi}(x'') \right) \right] \quad (6)$$

with $\Delta_x$ an elementary bond triangle with one corner being $x$ and the others $x'$ and $x''$. The coupling $\beta_2$ plays the role of $b_1$ in eq. (3). Again

---
[3] In the large $N$ limit, to simplify the calculation somewhat, slight variants where considered (see [11])

the large $N$ calculation shows that there is an optimal value for $\beta_2$ for which the bound on $M_H$ is largest. From now on this optimal choice for $\beta_2$ shall always be understood.

As already explained, to obtain a well defined bound on the Higgs mass we need to compute the cutoff effects on some physical quantity. We chose the cutoff effect in the square of the invariant $\pi-\pi$ scattering amplitude at $90^o$, shown in Figure 1.

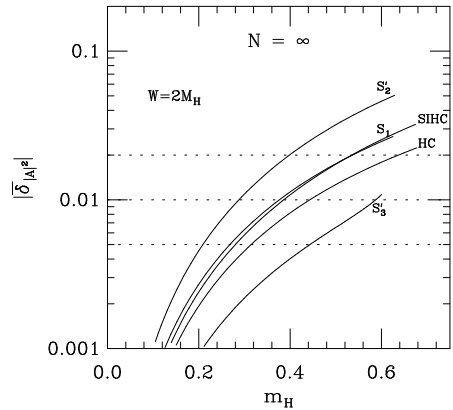

Figure 1. Leading order cutoff effects in the invariant $\pi - \pi$ scattering amplitude at $90^o$ at center of mass energy $W = 2M_H$ vs. the Higgs mass $m_H = aM_H$ in lattice units for the three $F_4$ actions, a hypercubic and a Symanzik improved hypercubic action. The three horizontal lines are at $\bar{\delta}_{|A|^2} = 0.005, 0.01, 0.02$.

### 2.3. The physical case $N = 4$

For the physical case, $N = 4$, one has to resort to numerical simulations. One needs measurements of $f = aF$, the pion decay constant, to set the scale, $F = 246\ GeV$, and of the Higgs mass. The results of the numerical simulations for the models in eq. (5) [14] and the newer results for the hypercubic and Symanzik improved hypercubic actions [12,15] are plotted in Figure 2. With the estimate of cutoff effects from the large $N$ calculation they can be turned into a bound on the Higgs mass.

We conclude that the MSM will describe physics to an accuracy of a few percent up to energies of the order 2 to 4 times the Higgs mass, $M_H$, only if $M_H \leq 710 \pm 60~GeV$. The error quoted accounts for the statistical errors, shown in Figure 2, as well as the systematic uncertainty associated with the remaining regularization dependence. Since this bound includes the result of a systematic search in the space of dimension six operators we expect it to hold in the *continuum*. A Higgs particle of mass $710~GeV$ is expected to have a width between $180~GeV$ (the perturbative estimate) and $280~GeV$ (the large $N$ nonperturbative estimate). Thus, if the Higgs mass bound turns out to be saturated in nature, the Higgs would be quite strongly interacting.

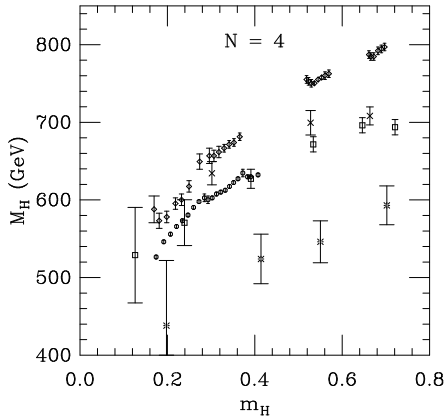

Figure 2. The Higgs mass $M_H = M_H/F \times 246~GeV$ in physical units vs. the Higgs mass $m_H = aM_H$ in lattice units from the numerical simulations. The bursts correspond to action $S_1$ [9], the squares to action $S_2$, the crosses to action $S_3$ [14], the octagons to the hypercubic and the diamonds to the Symanzik improved action [12].

### 2.4. Heavier Higgs with Ghosts?

All the work on the Higgs mass bound I have reviewed so far used a lattice regularization of some kind or other, which breaks the Lorentz invariance at least at order $1/\Lambda^4$ (hypercubic lattices break it at order $1/\Lambda^2$). An alternative, which avoids this, is to study higher derivative, or Pauli-Villars, regularizations. These, instead, have ghosts which lead to (and are) cutoff effects. But it might be that these effects are less severe than those of lattice regularizations. Large $N$ calculations indicate, that the cutoff effects may be slightly smaller than for the "best" lattice regularizations studied [11], but that the effect on the Higgs mass bound is small.

Jansen, Kuti and Liu [16] started to investigate the higher derivative model

$$S = \int_x \left[ \frac{1}{2} \vec{\phi}_c (-\partial^2 - \frac{1}{M^4}\partial^6) \vec{\phi}_c \right.$$
$$\left. -\frac{1}{2} m_0^2 \vec{\phi}_c^2 + \lambda_0 (\vec{\phi}_c^2)^2 \right] \quad (7)$$

with $M$ the regulator mass. To make numerical simulations possible, they put the model on a lattice, choosing $M < 1/a$ to keep the PV regulator scale $M$ below the lattice scale $1/a$, which is thought to be taken to infinity at the end.

The first preliminary numerical results came out rather surprising [16]: $m_H/f$ up to 8 was found with $m_H/M \approx 0.4$. This means that the Higgs mass could be as heavy as $2~TeV$, with the ghost states only appearing above $4~TeV$.

It will remain to be seen, whether this result holds up under closer scrutiny. In particular, one would like to know more about possible cutoff effects at lower energies. For example, it is possible to achieve $m_H/f \sim 8$ in a Nambu-Jona-Lasinio model [17], but it has an elementary fermion below the Higgs and hence strong cutoff effects.

It will be a challenge to estimate the cutoff effects for the Higgs model with ghosts. The result of [16], which is at least a factor two larger than large $N$ estimates, implies that the large $N$ expansion does not seem to work here. On the other hand, the corresponding renormalized coupling, $g_R \sim 200$, puts us far outside the regime of perturbation theory.

### 3. Fermion-Higgs Sector

Putting chiral fermions on a lattice is, at best, a nontrivial endeavor [18] and a method both satisfactory and numerically manageable is still missing. However, some ingredients of the Higgs-fermion sector of the MSM do not hinge on the chiral character of the fermions. Among these are the issue of triviality of Yukawa couplings, which like the scalar coupling are not asymptotically free, and the occurrence of a lower bound

on the Higgs mass, generally referred to as the "vacuum stability bound" [19].[4]

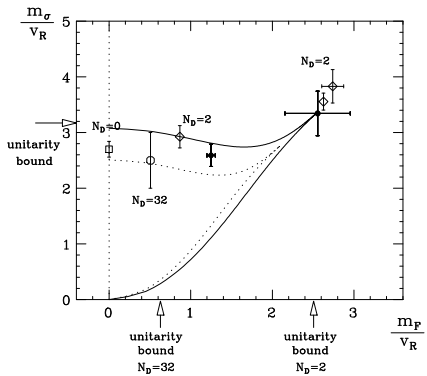

Figure 3. Upper bounds on $g_R$ as function of $y_R$ from [22]. Diamonds are infinite volume reduced staggered and full circle $6^3 \times 12$ mirror fermion results.

Compared to the studies in the pure scalar sector the bounds in Higgs-fermion models – the gauge couplings are still neglected and assumed to be treated perturbatively – are of more preliminary nature. For example, no study of cutoff effects in these models has been done to date. However, since the maximal renormalized scalar and Yukawa couplings found seem to lie in the region of validity of renormalized perturbation theory, I do not expect any surprises from a study of the cutoff effects.

The most recent results come from the study of reduced staggered fermions [20] and of mirror fermions [21]. For technical reasons, to obtain a positive definite fermion determinant that can be simulated with the Hybrid Monte Carlo algorithm, a doubling of the fermion content to two families is required for the numerical simulation in both approaches.

Sample results for the upper bound of the Higgs in the presence of heavy fermions are shown in Figure 3. The bound, slightly dependent on the fermion mass, is compatible with renormalized perturbation theory.

---
[4]On the lattice no sign of a vacuum instability has been observed. Rather the "vacuum stability bound" is the lower borderline of the region of renormalized couplings that can be obtained with allowed bare couplings, in particular with $\lambda \geq 0$.

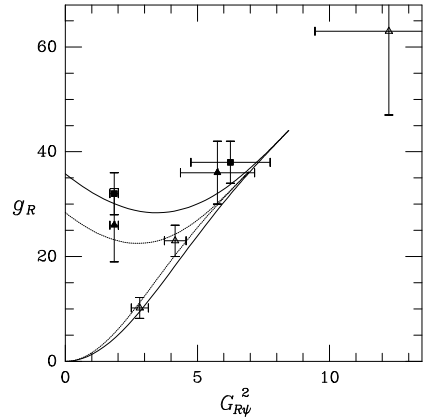

Figure 4. Upper and lower bounds on $g_R$ as function of $G^2_{R\psi}$ from a simulation with mirror fermions [25]. The solid and dotted lines are the perturbative estimates for scale ratios $\Lambda/m_R = 3$ and 4 ($\Lambda = \pi/a$).

I would also like to draw attention to the fact, illustrated in Figure 3, that no fermions much heavier than given by the tree-level unitarity bound for the Yukawa coupling, $\sqrt{4\pi/n_D}$, have been found. This seems to be a conclusions that holds for all Yukawa models studied so far, regardless of their symmetry group [23].

In the mirror fermion model both the upper [24] and lower [25] bound of the Higgs mass have been obtained in the exact decoupling case. They are shown in Figure 4. Again, the results appear compatible with renormalized perturbation theory.

## 4. Conclusions

Lattice calculations in the pure scalar sector with various regularizations, including actions with parameters allowing tuning of cutoff effects, have lead to the conclusion that $M_H \leq 710 \pm 60\, GeV$ if the MSM is to be a good description of physics up to energies 2 to 4 times $M_H$ to within a few percent.

Lattice calculations in Yukawa models, neglecting gauge interactions, have found bounds on Higgs and fermion masses that are compatible with renormalized perturbation theory. In particular no couplings greatly exceeding the tree-level unitarity bounds have been found.

A study of the pure scalar sector with a Pauli-Villars regulator lead to the surprising preliminary result that a Higgs bound of up to 2 $TeV$ without detectable ghost effects below $\approx 4$ $TeV$ seems possible. It is very important to check this result and in particular to carefully study the possible cutoff effects at lower energies.

**Acknowledgements**

I would like to thank M. Klomfass, P. Vranas and especially H. Neuberger for innumerable discussions during our collaboration on the subject of the Higgs mass bound. Thanks also go to W. Bock, C. Frick, J. Kuti, L. Lin, G. Münster, J. Smit and F. Zimmermann for correspondence and discussions while this review was prepared. This work was supported in part by the DOE under grants # DE-FG05-85ER250000 and # DE-FG05-92ER40742.